\documentclass[%
 reprint,
superscriptaddress,
 amsmath,amssymb,
 aps,
]{revtex4-2}

\usepackage[utf8]{inputenc}

\usepackage{graphicx}% Include figure files
\usepackage{dcolumn}% Align table columns on decimal point
\usepackage{bm}% bold math
\usepackage{color}

\begin{document}

\title{\normalsize Emission of particles from a parametrically driven condensate in a one-dimensional lattice}
\author{L. Q. Lai}
\affiliation{School of Physics and Electronics, Hunan University, Changsha 410082, China}
\affiliation{Laboratory of Atomic and Solid State Physics, Cornell University, Ithaca, New York 14853, USA}

\author{Y. B. Yu}
\affiliation{School of Physics and Electronics, Hunan University, Changsha 410082, China}

\author{Erich J. Mueller}
\email{em256@cornell.edu}
\affiliation{Laboratory of Atomic and Solid State Physics, Cornell University, Ithaca, New York 14853, USA}

\date{\today}

\begin{abstract}
Motivated by recent experiments, we calculate  particle  emission from a Bose-Einstein condensate trapped in a single deep well of a one-dimensional lattice when the interaction strength is modulated. In addition to pair emission, which has been widely studied, we observe single-particle emission. Within linear response, we are able to write closed-form expressions for the single-particle emission rates and reduce the pair emission rates to one-dimensional integrals. The full nonlinear theory of single-particle emission is reduced to a single variable integrodifferential equation, which we numerically solve.  
%We find a rich stability diagram, with competing instability channels.
%{\color{red}, leading to regimes of the excitations.}{\color{blue}(Appropriate?)}
\end{abstract}

\maketitle

\section{Introduction}

% Motivation
% \begin{itemize}
% \item Quantum dynamical system
% \begin{itemize}
% \item How does order develop in a quantum system?
% \item How do quantum correlations evolve?
% \item Relation to early universe:  Expansion, development of order, cosmic microwave background.
% \end{itemize}
% \item Cold atom experiment: fireworks
% \end{itemize}

Cold atom experiments have enabled previously unimaginable investigations of quantum dynamics, which combine the richness of classical dynamical systems with the profound and unexpected features of quantum mechanics. They explore fundamental questions of how order and correlations develop \cite{order1, order2}, and find extensive applications, including discovering novel nonequilibrium phases \cite{noneqphases1, noneqphases2} and modeling the evolution of the early universe \cite{universe1, universe2, universe3}. A recent experiment from the Chicago group \cite{fireworks1} and several follow-ups \cite{fireworks2, fireworks3, fireworks4, fireworks5, fireworks6, fireworks7} observed jets emerge from a gas of ultracold cesium atoms, when the interaction strength was modulated. This was both surprising, and visually striking.  Motivated by the phenomena, we develop and analyze a simple model of matter-wave emission, which reveals new aspects of such jet emission.  In particular, we find that in addition to the experimentally observed pair jets, there are regimes where one can see single-particle  emission.

The key technology behind these experiments is the ability to control the interaction strength of ultracold atoms \cite{chin, kengnea}.  At very low temperature, only $s$-wave collisions are allowed, and the low-energy scattering is quantified by a single number, the $s$-wave scattering length.  Magnetic fields mix in different scattering channels, and allow one to modify the scattering length.  Experiments have demonstrated both 
% That paper, and subsequent theore
% when they modulated the interaction strength in a gas of ultracold Cesium atoms
% a surprising and visually striking phenomenon of matter-wave jet-emission from and 
% The realization of tuning the interactions in ultracold atomic gases leads to remarkable phenomena, and enables a variety of new applications \cite{chin, kengnea}. Of particular interest in recent years is to 
temporal \cite{temporal1, temporal2, temporal3, temporal4} and spatial \cite{spatial1, spatial2, spatial3, spatial4} control of the interactions, enabling an incredibly wide range of 
explorations \cite{topological, mott, gauge, simulation, temporal3}.
%modulate the interaction strength near a Feshbach resonance to investigate how orders develop and how correlations evolve in these nonequilibrium quantum dynamical systems, which provides us versatile platforms to study, for instance, 
%gauge fields, Mott transitions, topological effects, and quantum simulations 
%, and also can be utilized as seminal analog systems for inflationary early Universe and cosmological expansion \cite{universe1, universe2, universe3}.
In the jet experiments \cite{fireworks1}  a spatially uniform magnetic field is sinusoidally modulated at frequencies $\omega/2\pi \sim$ kHz -- which are fast compared to typical timescales of collective oscillations, but very slow compared to atomic excitations.  As a consequence, the scattering length oscillates, which leads to particle jets.

This experiment has been modeled using Bogoliubov theory \cite{fireworks1,yan, zhai, holland}.   The oscillating scattering length appears as a parametric drive in the equations for the elementary excitations of the condensate.  The drive resonantly excites pairs of particles, each of which has energy $\hbar\omega/2$, and these form the jets. 
The quantum state, with these strong pair correlations, is quite exotic. At larger drive strength they also observed nonlinear processes, where the outcoming particles have multiples of this energy \cite{fireworks4}.

% Boson double state?
%{\color{red} 

%, as well as analyzing the full nonlinear Gross-Pittaevskii equations.  We find the instability to pair-jets, as 
%argue that one can also produce single-particle jets, where the resulting particles gain energy $n \hbar\omega$, for integer $n$. These single-particle emission processes are distinct from the Bogoliubov excitations seen in the experiments.
%{\color{blue} (Appropriate?)}
%(We also find higher-order processes which impart energy $n\hbar\omega$, for integer $n$.)  
%We introduce a simple model for exploring these non-linear effects.  
%These single-particle jets would be observable in an experiment with either lower oscillation frequencies or deeper traps.

We analyze such particle emission within a 1D lattice model. This  experimentally accessible geometry is chosen to make the analysis as simple as possible, thereby making the phenomena as clear as possible.  This model has a finite bandwidth, which spectrally separates various processes.  
%We believe that the qualitative features should be universal, and the same physics can be seen in other geometries.  
We present two approaches:  linear response theory and mean-field theory. 

In the linear regime we use Green's functions to calculate the emission rates. We find two distinct instabilities:  The pair emissions observed in the experiment, and a distinct single-particle emission process. 
To explore the full nonlinear behavior, we convert the lattice Gross-Pittaevskii equation into a single variable nonlinear integrodifferential equation:  effectively a damped nonlinear oscillator with a non-Markovian bath.  We numerically integrate these equations, and find a series of higher-order single-particle emission processes.

Related physics is seen in amplitude or phase modulated lattices \cite{pedersen,cabrera,arnal,wintersperger,kramer,stoferle,gemelke}.  The primary difference being that the case of modulated interactions is intrinsically nonlinear.  Modeling of these amplitude modulated lattices has largely focused on the harmonically confined system, where any jets that are formed remain trapped \cite{das,yamakoshi,yamakoshi2}.
Parametric excitation of condensates has also been explored in a number of other contexts \cite{bucker,wasak,bonneau}.

In Sec.~\ref{model} we introduce our model. In Sec.~\ref{meanfield} we describe the mean-field theory approach. In Sec.~\ref{linear} we use linear response theory to
calculate the single-particle and pair emission rates.  In Sec.~\ref{dynamics} we present numerical analysis of the mean-field theory from Sec.~\ref{meanfield}. We summarize our results and their implications in Sec.~\ref{summary}.

\section{Model}
\label{model}
%In the Chicago experiment, a condensate is confined vertically in a finite-depth puck-shape trap, and the interaction strength is modulated at angular frequency $\omega$.  They observe particle jets flying out in the horizontal plane, which is perpendicular to the trap's axis.

\begin{figure}[tbp]
\includegraphics[width=\columnwidth]{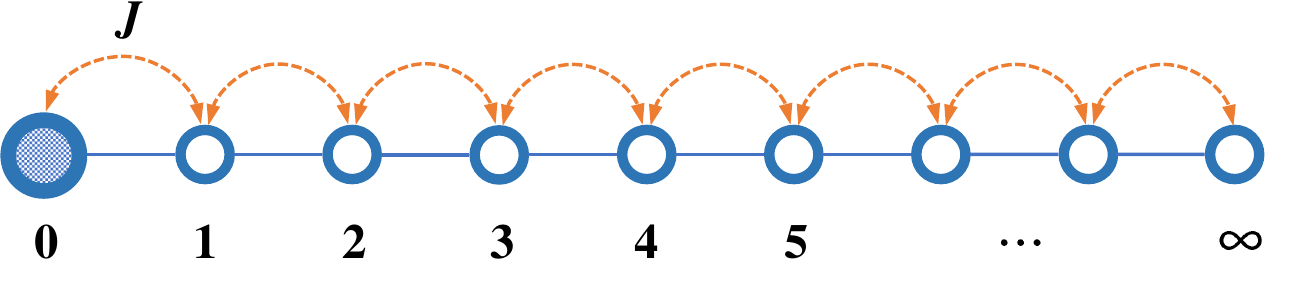}
\caption{Schematic of the 1D semi-infinite lattice. A potential, of depth $V$ is applied to  site $0$, and the pairwise interaction energy of atoms on that site has strength $U$.  Hopping between neighboring sites has matrix element $J$.}
\label{latticepic}
\end{figure}

%There are many interesting facets of their observations.  Here we introduce a minimal model which lets us explore one aspect.  
We consider a 1D semi-infinite lattice, as depicted in Fig.~\ref{latticepic}.  The sites are labeled by integers $j$, running from 0 to $\infty$, and the site at $j=0$ represents the trap.  We apply a local potential at that site to confine the atoms.  Atoms which escape can hop down the chain, and move off to infinity.

We choose this semi-infinite geometry to eliminate as many complications as possible.  By having only a single ``lead,"  all jets must propagate in that one direction, and we do not need to consider correlations between jets that move in different directions.  One can readily engineer this geometry in an experiment.

As a further attempt to keep the model simple, we only include interactions between atoms which sit on the site $j=0$.  The physical justification is that the atomic density will be low outside of the trap, and it is reasonable to neglect interactions in that region. Inhomogeneous magnetic fields, and a Feshbach resonance, can be used to literally implement a model with such  spatially localized interactions.
%This spatially localized interaction can be 
%If one wants to literally implement our model, one can use spatially dependent magnetic fields, and atoms near Feshbach resonances, to fulfill such a spatially localized interaction.  
%Such an experiment would be challenging.

Mathematically, our model is described by the Hamiltonian
\begin{eqnarray}
\hat{H} &=&V \hat{a}_{0}^{\dag }\hat{a}_{0}+\frac{1}{2}
\left( U+ g\left( t\right) \right) \hat{a}_{0}^{\dag }\hat{a}_{0}^{\dag }\hat{a}_{0}\hat{a}_{0}  \nonumber \\
&&-J\sum_{j=0}^{\infty }\left( \hat{a}_{j+1}^{\dag }\hat{a}_{j}+\hat{a}_{j}^{\dag
}\hat{a}_{j+1}\right).
\end{eqnarray}
Here, $V<0$ is the trapping potential, $\hat{a}_j^{\dag}$ ($\hat{a}_j$) are creation (annihilation) operators, and $J$ quantifies the hopping between nearest-neighbour sites. The time-dependent interactions are characterized by a constant term $U$, and a sinusoidally oscillating term $g\left(t\right)=g\sin\left(\omega t\right)$ or $g\left(t\right)=g\sin\left(\omega t\right) \theta(t)$, where $g$ is the drive strength. The step function $\theta(t)$ is included to model the situation where the oscillations are suddenly turned on.

\section{Mean-Field Equations} \label{meanfield}
We begin by constructing the mean-field equations of motion, and finding the steady state solution when $g=0$.  
%This will be the starting point for all of our subsequent calculations.
We replace
the operators with their expectation values $a_j=\langle \hat a_j\rangle$.  
%To describe the pair emission in
% Ref.~\cite{fireworks1}, 
% we would need to supplement these Gross-Pittaevskii equations with the Bogoliubov equations for the quadratic fluctuations.  We choose not to include these fluctuations: There are many excellent treatments of this pair-creation
% which treat the excitations through a Bogoliubov approximation 
%\cite{yan, zhai, holland}, and all of our novel physics involves the mean-field equations. 
% In Section~\ref{} we discuss the relations between these approaches.
%Although there are a number of important subtleties, the Bogoliubov approach can be interpreted as a linearization of the mean-field theory \cite{pethick}.  
Physically, $|a_j|^2$ corresponds to the number of particles on site $j$, and $I_j=2J\,{\rm Im}(a_{j+1}^*a_j)$ is the particle current flowing from site $j$ to $j+1$.  

On site $j=0$ the expectation value of the  Heisenberg equations of motion read ($\hbar=1$ throughout this paper)
\begin{eqnarray}
i\partial _{t}a_{0} &=&\langle \left[ \hat a_{0},H\right]\rangle   \nonumber \\
&=& V a_{0}+\left(U+g(t)\right) a_{0}^*a_{0}a_{0}-Ja_{1},\label{mf0}
\end{eqnarray}
where we have neglected fluctuation terms.  In Sec.~\ref{linear} we reintroduce these fluctuations, and argue that they play no role unless the parametric drive is resonant with pair emission processes.  The parameters of our model can be chosen so that pair emission and single-particle emission are spectrally isolated, and can be treated independently.

On the remaining sites, where $j>0$,
\begin{equation} \label{rest}
i\partial_t a_j(t) = -J \left(a_{j+1}(t)+a_{j-1}(t)\right).
\end{equation}
Because we have neglected interactions on these sites, this latter set of equations is linear. We can formally solve Eq.~(\ref{rest}) under the assumption that there are no particles entering the system from infinitely far away,
\begin{equation} \label{ajt}
a_j(t) = -J \int^td\tau\,G_{j1}(t-\tau) a_0(\tau),
\end{equation}
where as derived in Appendix~\ref{Green's function},
the Green's function is
\begin{equation}
G_{j1}(t)= i^{j-2} \frac{j J_j(2 J t)}{J t} \theta(t).
\end{equation}
Here, $J_{n}\left(z\right)$ is the Bessel function of the first kind.
%This formal solution allows us to eliminate all sites with $j>0$, resulting 
We thereby arrive at 
a nonlinear integrodifferential equation
\begin{equation}\label{mastereqn}
i\partial_t a_0 = V a_0 + (U+g(t)) a_0^* a_0 a_0 + J^2 \int^t d\tau G_{11}(t-\tau) a_0(\tau).
\end{equation}
We find the stationary solution by making the ansatz $a_0=\alpha e^{-i\nu t}$, whence
\begin{equation}\label{stat}
\nu=V+U|\alpha|^2+ J^2 G_{11}(\nu),
\end{equation}
where, as derived in Appendix~\ref{Green's function}, the frequency-domain Green's function is
\begin{equation}
J G_{11}(\epsilon)=\frac{\epsilon}{2 J}-i \sqrt{1-\frac{\epsilon^2}{4J^2}}.
\end{equation}
%To produce the correct branch one should add a small positive imaginary part to $\nu$ and take the principle branch of the square root.  

Equation~(\ref{stat}) is solved by isolating the square root on one side of the equation, and squaring both sides.  The resulting linear equation gives
\begin{equation}
\nu = \frac{J^2 + \left(V+U |\alpha|^2\right)^2}{V+U|\alpha|^2}.
\end{equation}
Substituting back into the original equation, we find that this is a spurious root if 
$|V+U|\alpha|^2|<J$.  Under those conditions, the trap is unable to contain the particles.

The case where 
$V+U \left\vert \alpha \right\vert^2 <-J$ corresponds to a conventional bound state sitting below the continuum, while for $ V+U \left\vert \alpha \right\vert^2 >J$ it is a ``repulsively bound state" sitting above the continuum.  The latter exists because the spectrum is bounded.  

Note, the same results can be found by substituting the ansatz $a_j=\alpha e^{-i\nu t} e^{-\kappa j}$ into Eqs.~(\ref{mf0}) and (\ref{rest}).

\section{Linear Response}\label{linear}
Here we calculate the rate of particle emission when $g$ is small.
We start from the ansatz
\begin{equation}\label{gauge}
\hat a_j(t)=e^{-i\nu t}\left(\alpha_j +\hat b_j(t)\right),
\end{equation}
where $\alpha_j$ are the solutions to the mean-field equations with $g=0$:  Eqs.~(\ref{mf0}) and (\ref{rest}).  To ease notation, we leave off the index $j$ when $j=0$, i.e., we define $\alpha=\alpha_0$, and take $\alpha$ to be real.    
%The expectation value of $\hat b_j$ vanishes in the unperturbed state,
%$\langle \hat b_j \rangle_0=0$, but not necessarily in the unperturbed state.  
% Thus, up to quadratic fluctuations, the number of condensed particles on site $j=0$ is
% $N_c=|\alpha|^2 +\alpha^* 
% \langle \hat b_0\rangle+ \alpha \langle\hat b_0^\dagger\rangle$.

\begin{figure}[tbph]
\includegraphics[width=\columnwidth]{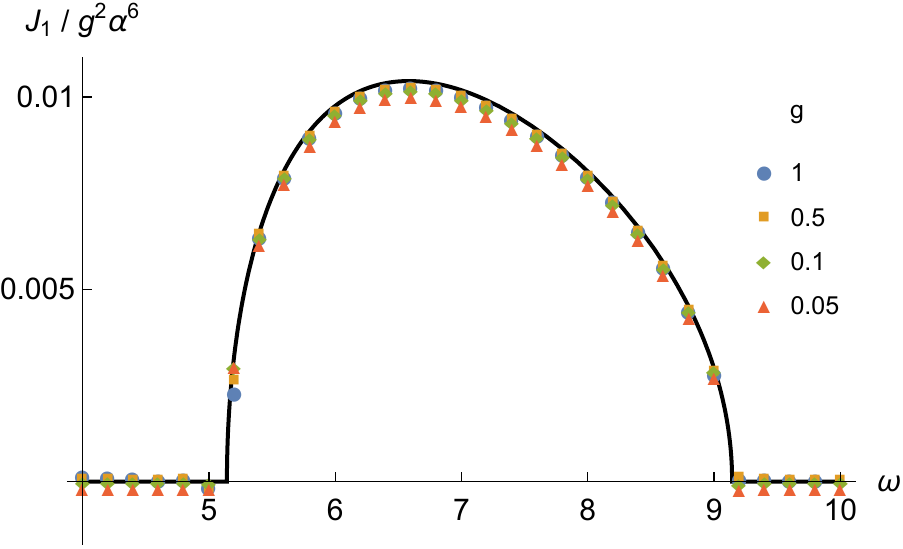}
\caption{Single-particle emission rate $J_1=\langle \hat J_1\rangle$ calculated from  linear response, Eq.~(\ref{j1lin}) (solid curve), and by fitting to simulations of  the nonlinear mean-field equations, in Sec.~\ref{dynamics}.  Here we have used units where $J=1$, and taken $V=-7$, $|a_{0}(t=0)|^2=1$, and $U=0$, so that $\nu=-7.14$.  Linear response theory works well in this regime, as is evident by the collapse of the data with different drive strength $g$.  The small discrepancy is an artifact related to extracting the emission rate from numerical data. }
\label{compare}
\end{figure}

Equation~(\ref{gauge}) is a canonical transformation, and up to quadratic order the transformed Hamiltonian is
\begin{eqnarray}
\hat H&=& E_0 + \hat{H}_{\rm bog}+ g(t)\left(\hat X_1 + \hat X_2\right),
\end{eqnarray}
with
\begin{eqnarray}
\hat H_{\rm bog}&=& 
(V+2U\alpha^2 -\nu)\hat b_0^\dagger b_0 + \frac{U\alpha^2}{2} \left(\hat b_0 \hat b_0
+ \hat b_0^\dagger \hat b_0^\dagger\right) \nonumber\\
&&+\sum_{j=1}^\infty \left(-\nu \hat b_j^\dagger \hat b_j - J \hat b_j^\dagger \hat b_{j-1} - J \hat b_{j-1}^\dagger \hat b_j\right), \\
\hat X_1 &=&  \alpha^3 \left(\hat b_0 + \hat b_0^\dagger\right), \\
\hat X_2 &=&
\frac{\alpha^2}{2}\left(
 \hat b_0 \hat b_0 +
 \hat b_0^\dagger \hat b_0^\dagger
+ 4\hat b_0^\dagger \hat b_0
\right),
\end{eqnarray}
where $E_0$ is the ground-state energy and $\hat H_{\rm bog}$ describes the quadratic fluctuations about the ground state.  The perturbation is broken into two terms. At linear order, $\hat X_1$ is responsible for single-particle emission.  The operator whose expectation value give the  emission rate is 
\begin{eqnarray}
\hat J_1&=&\frac{1}{i}[\hat b_0^\dagger \hat b_0 ,g(t) \hat X_1 ]=  g(t) \alpha^3  \frac{\hat b_0^\dagger- \hat b_0}{i}.
\end{eqnarray}
Similarly, the operator corresponding to the particle flux from pair emission is
\begin{eqnarray}
\hat J_2&=&\frac{1}{i}[\hat b_0^\dagger \hat b_0,g(t) \hat X_2 ]=g(t) \alpha^2\frac{
\hat b_0^\dagger \hat b_0^\dagger
- \hat b_0 \hat b_0}{i}.
\end{eqnarray}
Within linear response the single-particle and pair processes can be treated separately.  It is convenient to write $\hat J_1 = -i g(t) \hat Y_1   $ and $\hat J_2 = -i g(t) \hat Y_2$.  Time-dependent perturbation theory then gives
\begin{eqnarray}
\langle \hat J_j\rangle &=&
-\int^t d\tau g(t)g(\tau) \langle[\hat Y_j(t),\hat X_j(\tau)] \rangle,
\end{eqnarray}
where the expectation value is taken in the $g=0$ state, and $\hat {\cal O}(t)= e^{i \hat H_{\rm bog} t} \hat {\cal O} e^{-i \hat H_{\rm bog} t}.$
We take $g(t)=g\sin(\omega t)$, and average over one period to get
\begin{equation}
\langle \hat J_j\rangle 
=\frac{g^2}{4}{\rm Im}\left(
\chi_j(\omega)+\chi_j(-\omega)
\right),
\end{equation}
where $\chi_j (t-t^\prime)=-i\theta(t-t^\prime) \langle [\hat Y_j(t),\hat X_j(t^\prime)]\rangle$ is the retarded response function, and we have used that both $\hat X$ and $i \hat Y$ are Hermitian to write $\chi^*(t)=-\chi(t)$ and hence $\chi(\omega)=-\chi^*(-\omega)$.  

We now write these response functions in terms of the $2\times 2$ matrix Green's function
\begin{equation}
{\cal G}_{ij}(t-t^\prime) = \frac{1}{i}\theta(t-t^\prime) \langle [\hat \phi_i(t),\hat \phi_j^\dagger(t^\prime)]\rangle,
\end{equation}
where $\hat \phi_1=\hat b_0$ and $\hat \phi_2=\hat b_0^\dagger$.   Fourier transforming the equations of motion for $\cal G$ gives
\begin{equation}\label{eom}
\left(\begin{array}{cc}
\omega+\nu -\Delta_{\omega+\nu}& -\alpha^2 U\\
-\alpha^2 U& -\omega+\nu-\Delta^*_{-\omega+\nu}
\end{array}\right){\cal G}=
\left(
\begin{array}{cc}
1&0\\
0&1
\end{array}
\right)
\end{equation}
with
\begin{equation}
\Delta_\epsilon=V+2 \alpha^2 U+J^2 G_{11}(\epsilon).
\end{equation} 
The inversion is straightforward, and ${\cal G}(\omega)$ is real unless $-2J<\omega\pm\nu<2J$, and hence $\langle\hat J_1\rangle$ vanishes outside that region.
Formally,
\begin{equation}\label{j1lin}
\langle \hat J_1\rangle
 =
  \frac{g^2 \alpha^6}{4 i}
  \left(\begin{array}{cc}-1&1\end{array}\right) \left({\cal G}(\omega)+{\cal G}(-\omega)\right) \left(
  \begin{array}{cc}
  1\\
  1
  \end{array}
  \right).
%  &=&\langle \hat J_1\rangle=
%  \frac{g^2 J^2 |\alpha|^6}{i}\left[G_{11}(\omega+\nu)-G_{11}(-\omega-\nu)+G_{11}(-\omega+\nu)-G_{11}(\omega-\nu)\right]\\
 %&=& g^2  |\alpha|^6 \sqrt{4 %J^2-(\omega\pm\nu)^2}.
\end{equation}
In Appendix~\ref{multiscale}, we give a more elementary derivation of this result, which demonstrates that this result is captured by mean-field theory.

\begin{figure}[tbp]
\includegraphics[width=\columnwidth]{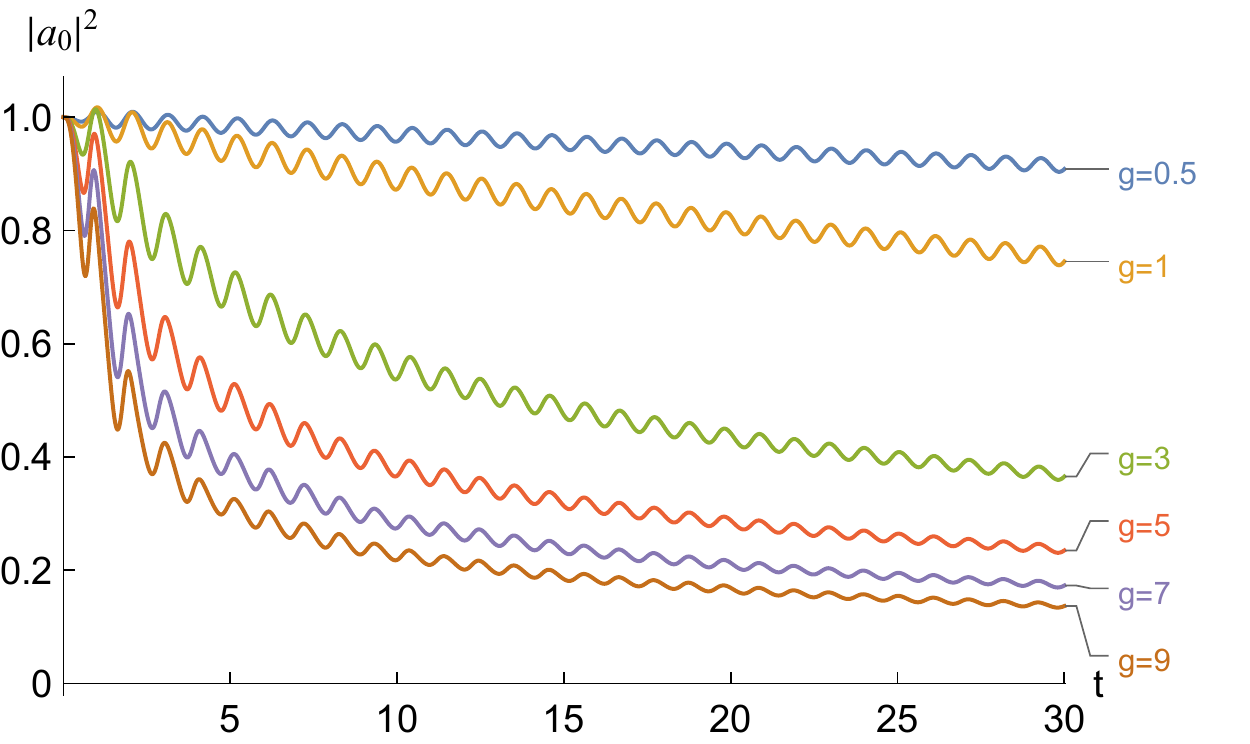}
\caption{Time dependence of the condensate for different drive strength $g$ with fixed interaction strength $U=1$. The trapping potential is $V=-7$, and the drive frequency is $\omega=6$. Energies are in units of $J$, and times are in units of $\hbar/J$.}
\label{gdecaypic}
\end{figure}

%This same result can be calculated by more elementary means -- see Appendix~\ref{multiscale}.

Figure~\ref{compare} shows a typical emission spectrum. As expected, it has finite support. Also shown are emission rates calculated via the techniques in Sec.~\ref{dynamics}.  These should agree when $g$ is small, where linear response theory is applicable.

\begin{figure*}[tbp]
\includegraphics[width=\textwidth]{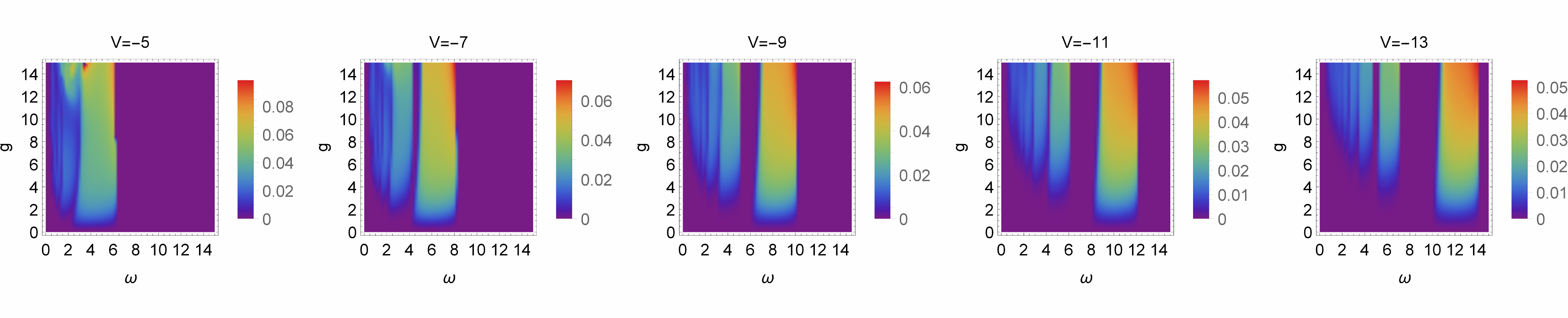}
\centering
\caption{(Color Online) Condensate decay rate $\Gamma$ vs. drive strength $g$ and drive frequency $\omega$ for different trapping strength $V=-5,-7,-9,-11,-13$. Color bars denote $\Gamma$. In all cases the interaction strength is fixed as $U=1$.   The chemical potentials are
$\nu=-4.25,-6.17,-8.13,-10.10,-12.08$, respectively. All energies are in units of $J$.}
\label{gdecayratepic}
\end{figure*}

Calculating the pair emission rate is more difficult, as it involves correlations of four operators.  It is convenient to introduce functions
${\cal G}^>_{ij}=\langle \hat \phi_i(t) \phi_j^\dagger(0)\rangle$ and ${\cal G}^<_{ij}=\langle  \phi_j^\dagger(0)\hat \phi_i(t)\rangle$.
Since $\hat H_{\rm bog}$ is quadratic, we can apply Wick's theorem, and write
$\chi(t)= -i \theta(t) [\chi^>(t)-\chi^<(t)]$ with
\begin{eqnarray}\label{pairchi}
\chi^>(t)&=& \alpha^4 \left({\cal G}^>_{22}{\cal G}^>_{22}+{\cal G}^>_{21}{\cal G}^>_{21} +4 {\cal G}^>_{22}{\cal G}^>_{21}\right.\nonumber\\
&&\quad \left. -{\cal G}^>_{11}{\cal G}^>_{11}-{\cal G}^>_{12}{\cal G}^>_{12} -4 {\cal G}^>_{11}{\cal G}^>_{12}
\right),
\end{eqnarray}
and a similar expression for $\chi^<(t)$.  Since we are working in the Bogoliubov vacuum, the Fourier transform gives ${\cal G}^>(\omega)=A(\omega) \theta(\omega)$ and
${\cal G}^<(\omega)=-A(\omega) \theta(-\omega)$, where $A(\omega)=2{\rm Im} {\cal G}(\omega)$ is the spectral density \cite{pethick}. We then note that $\Pi(\omega)=2{\rm Im} \chi(\omega) = \chi^>(\omega)-\chi^<(\omega)$  to express $\langle \hat J_2\rangle$ as a convolution between elements of $A_{ij}(\omega)$ times step functions.  Since the spectral densities have finite support, $\langle \hat J_2\rangle$ vanishes unless $-2J<\omega/2 \pm \nu<+2 J$.

In Appendix~\ref{U=0}, we give explicit expressions for both $\langle \hat J_1\rangle$ and $\langle\hat J_2\rangle$ in the limit $U=0$.

\section{Nonlinear Dynamics}
\label{dynamics}

We now consider the full nonlinear behavior. We work at the mean-field level, largely to simplify the numerics.  %because including fluctuations beyond linear response is very challenging.  
In our context, this treatment cannot capture the physics of pair emission, but it does describe single-particle emission. This approximation should be valid when the pair and single-particle excitations are spectrally separated.
Note, that when the condensate has more degrees of freedom, such a mean-field approach will be able to capture the physics of pair jets \cite{fireworks2}.  Dynamical broken symmetries couple the different emission channels.

%with $g(t)=g \sin(\omega t)\theta(t)$.  That is, 
We imagine that the perturbation is turned on at time $t=0$, and 
take $g(t)=g \sin(\omega t)\theta(t)$.
We assume that the system is in equilibrium for $t<0$, and therefore take 
%To numerically integrate the integro-differential equation in
% Eq.~(\ref{mastereqn}), we use the condition
$a_0(t<0)=\alpha e^{-i\nu t}$.  This allows us to write Eq.~(\ref{mastereqn}) as
\begin{eqnarray} 
i\partial _{t}a_{0}(t)  &=& V a_{0}(t) +\left[U+g(t)\right]  a_{0}^{\ast}(t) a_{0}(t)
a_{0}(t)  \nonumber \\
&&+J^{2}\int_{0}^{t} G_{11}(t-\tau) \left[a_{0}(\tau)-\alpha e^{-i\nu \tau}\right]d\tau \nonumber \\
&&+\alpha e^{-i\nu t}J^{2}G_{11}(\nu).
\end{eqnarray}
We choose a fixed time step, using the Runge-Kutta (RK4) method to evolve $a_0$. In evaluating the right-hand side of the integrodifferential equation, we use the $a_0$'s from prior time steps, and calculate the integral utilizing Simpson's rule \cite{simpson}. We repeat the calculations for multiple step sizes to verify that the finite step-size error is negligible.

% We numerically integrate Eq.~(\ref{mastereqn}) using an approach detailed in Appendix~\ref{approach}.  For $t<0$, we take $a_0(t)$ to be given by the stationary solution when $g=0$,
% \begin{equation} \label{stationary}
% a_0(t)=\alpha e^{-i\nu t}.
% \end{equation}
% As shown in Appendix~\ref{nu}, such stationary solutions can be found if
% $| V+U \left\vert \alpha \right\vert^2 |>J$, and correspond to a bound state with
% \begin{eqnarray}
% \nu =\frac{J^2+\left( V+U\left\vert \alpha \right\vert ^{2}\right) ^{2}}{%
% \left( V+U\left\vert \alpha \right\vert ^{2}\right) }.
% \end{eqnarray}
% If $| V+U \left\vert \alpha \right\vert^2 |<J$, there is no stationary solution, as the well is too shallow. The case where 
% $V+U \left\vert \alpha \right\vert^2 <-J$ corresponds to a conventional bound state sitting below the continuum, while for $ V+U \left\vert \alpha \right\vert^2 >J$ it is a ``repulsively bound state" sitting above the continuum.  The repulsively bound state exists because the spectrum is bounded.  

Without any loss of generality we take $\alpha=1$, which is accomplished by scaling $a_j\to a_j/\alpha$,  $U\to U|\alpha|^2$ and $g\to g |\alpha|^2$. For most of our numerical analysis, we will work in units where $J=1$ -- though we reintroduce the scale $J$ in our discussions as appropriate.
%for some of our analytic calculations it will be convenient to explicitly include $J$.  

Figure~\ref{gdecaypic} represents typical values for the time dependence of the number of condensed particles in the well, after the oscillating modulation is turned on. We see two different behaviors:  For some parameters the number simply oscillates. This corresponds to the situation where the condensate wavefunction undergoes oscillations, but the atoms remain bound. For other parameters the number falls.
%, and approaches a saturated value after a sufficient modulating time.
This latter case corresponds to  particle emission, similar to what is seen in Ref.~\cite{fireworks1}. % As already explained, here we are seeing single-particle emission, rather than pair emission.

The condensate decay is nonexponential. The arguments from  Appendix~\ref{multiscale} imply that for small $g$ the decay rate is
\begin{equation}\label{ad}
\partial_t |a_0(t)|^2=-\langle \hat J_1\rangle,
\end{equation}
where $\langle \hat J_1\rangle$ is given by Eq.~(\ref{j1lin}) with $\alpha$ replaced by $a_0(t)$.  The right-hand side is a highly nonlinear function of $|a_0(t)|^2$, especially when $U$ is large.  Moreover, Eq.~(\ref{ad}) will break down when $g$ is large.
%for  
 % For short times, however, one can define a decay rate $\Gamma=J_1/|\alpha|^2$.
Nonetheless, we quantify the decay by fitting the condensate number to an exponential, $|a_0(t)|^2= A e^{-\Gamma t}$. In this fit, $\Gamma$ corresponds to the average rate at which particles are emitted from the condensate, and in the linear regime $\Gamma=\langle \hat J_1 \rangle/|\alpha|^2$.  Figure~\ref{compare} verifies that for small $g$ we reproduce the linear response results. 

Due to the nonexponential nature of the decay, our calculated $\Gamma$ has a weak dependence on the simulation time. For the comparison in Fig.~\ref{compare}, where numerical precision is important, we extrapolate to the short time limit.  In all further graphs, where we are mainly interested in qualitative features, we perform a single fit over $0<t<30/J$.

Figure~\ref{gdecayratepic} shows how this rate depends on the drive frequency $\omega$ and the drive strength $g$ when the well depth $V$ ranges from relatively shallow to deep.  We take the interaction strength $U=1$ -- though similar results can be found for different $U$. For generic drive frequencies and weak drive strength, the condensate is stable and $\Gamma=0$. We see that for $|\nu|-2J<\omega<|\nu|+2J$, even a small $g$ leads to particle emission.  This restriction is related to energy conservation. Generally, a particle in the condensate has energy $\nu$, while the particle which escapes to infinity must have $-2J<E<2J$. Although not shown here, we find quantitative agreement with the linearized model in Sec.~\ref{linear}.
%In Appendix~\ref{comparison} we perturbatively analyze Eq.~(\ref{mastereqn}), and establish this stability criterion.

One can see further bands of instability at finite drive strength $g$ when $|\nu|-2J<n \omega<|\nu|+2J$, for integer $n$.  These correspond to nonlinear processes, and are suppressed at small $g$. When the well is shallow (exemplified by $V=-5$), the different instabilities overlap, while for deeper wells (e.g., $V=-13$) they would be separated from one-another. Importantly, the linearized model cannot capture this physics.

%Figure~\ref{udecayratepic} corresponds to some of these higher-order resonances.

% If we included the Bogoliubov excitations, we would find another window of instabilities, when 
% $|\nu|-2J< \omega/2<|\nu|+2J$.  Outside this window, these excitations should have negligable effect.

\begin{figure}[bp]
\includegraphics[width=\columnwidth]{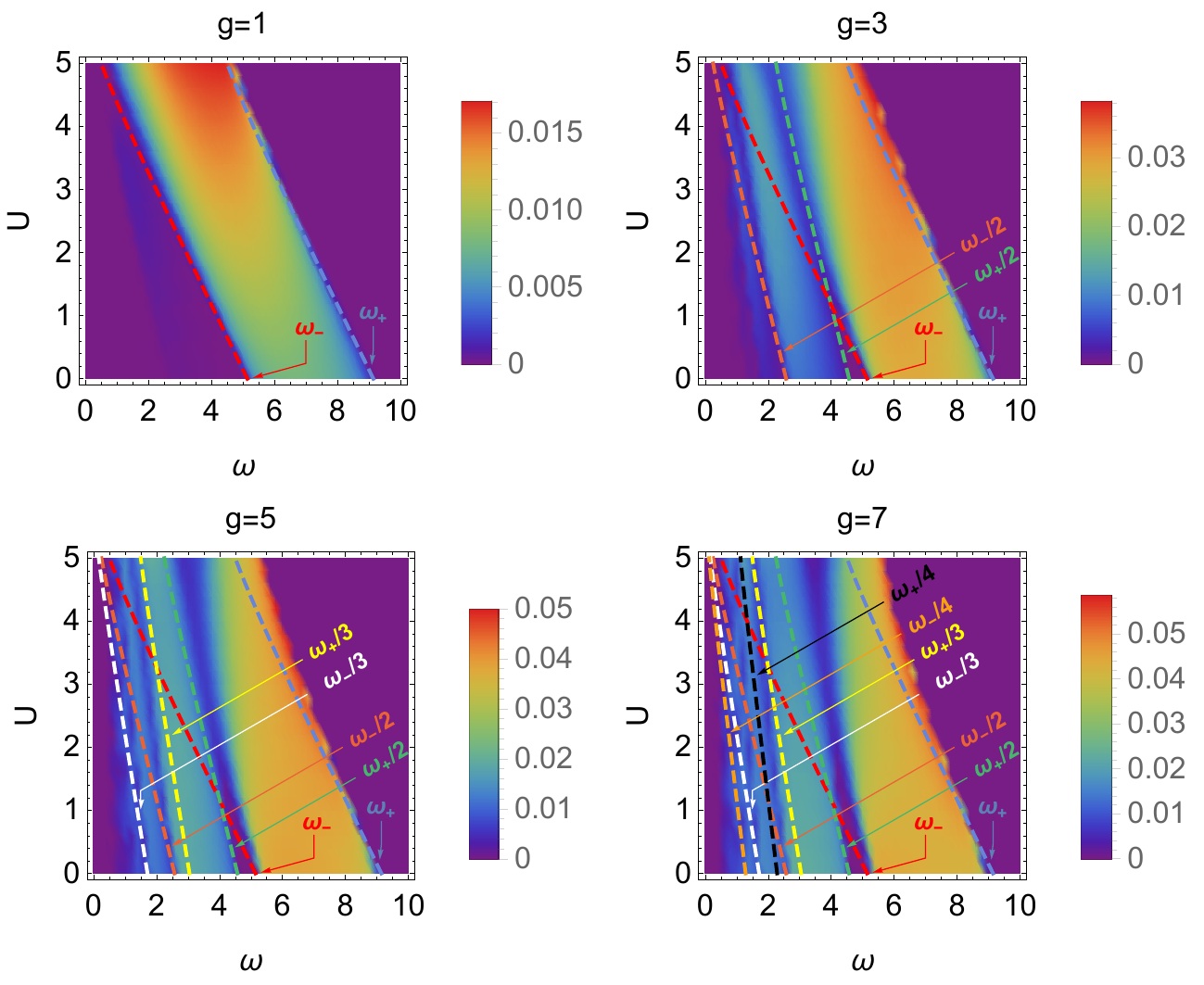}
\caption{(Color online) Condensate decay rate $\Gamma$ vs. interaction strength $U$ and drive frequency $\omega$ for different drive strength $g$. Color bars denote $\Gamma$.  The trapping potential is fixed as $V=-7$. Also shown are characteristic scales $\omega_{\pm}/n=(|\nu|\pm 2J)/n$, for integer $n$.   These delineate the regimes where $n$'th order excitations would be expected at weak coupling. All energies are in units of $J$.  }
\label{udecayratepic}
\end{figure}

Figure~\ref{udecayratepic} demonstrates how the instability depends on the interaction strength $U$ and the drive frequency $\omega$, for a moderately deep well $V=-7$.  When the drive strength is moderately weak ($g=1$), one only sees the most dominant instability, delimited by $\omega_-<\omega_{1}<\omega_+$ with $\omega_\pm = |\nu|\pm 2J$. The nonlinear excitations become prominent when $g$ is larger.  For example, one sees the second-order excitation at $\omega_-<2\omega_{2}<\omega_+$ when $g=3$. For $g=5$ and $g=7$, there are also third-order excitation at $\omega_-<3\omega_{3}<\omega_+$ and fourth-order excitation at $\omega_-<4\omega_{4}<\omega_+$.  These regions overlap, leading to a complicated pattern.   For large $g$ (for example, $g=7$) the higher-order effects appear to also renormalize the locations of the boundaries.
%, in which the frequency regime are narrower, and have overlapping with the first- and second-order ones.

\begin{figure}[tbp]
\includegraphics[width=\columnwidth]{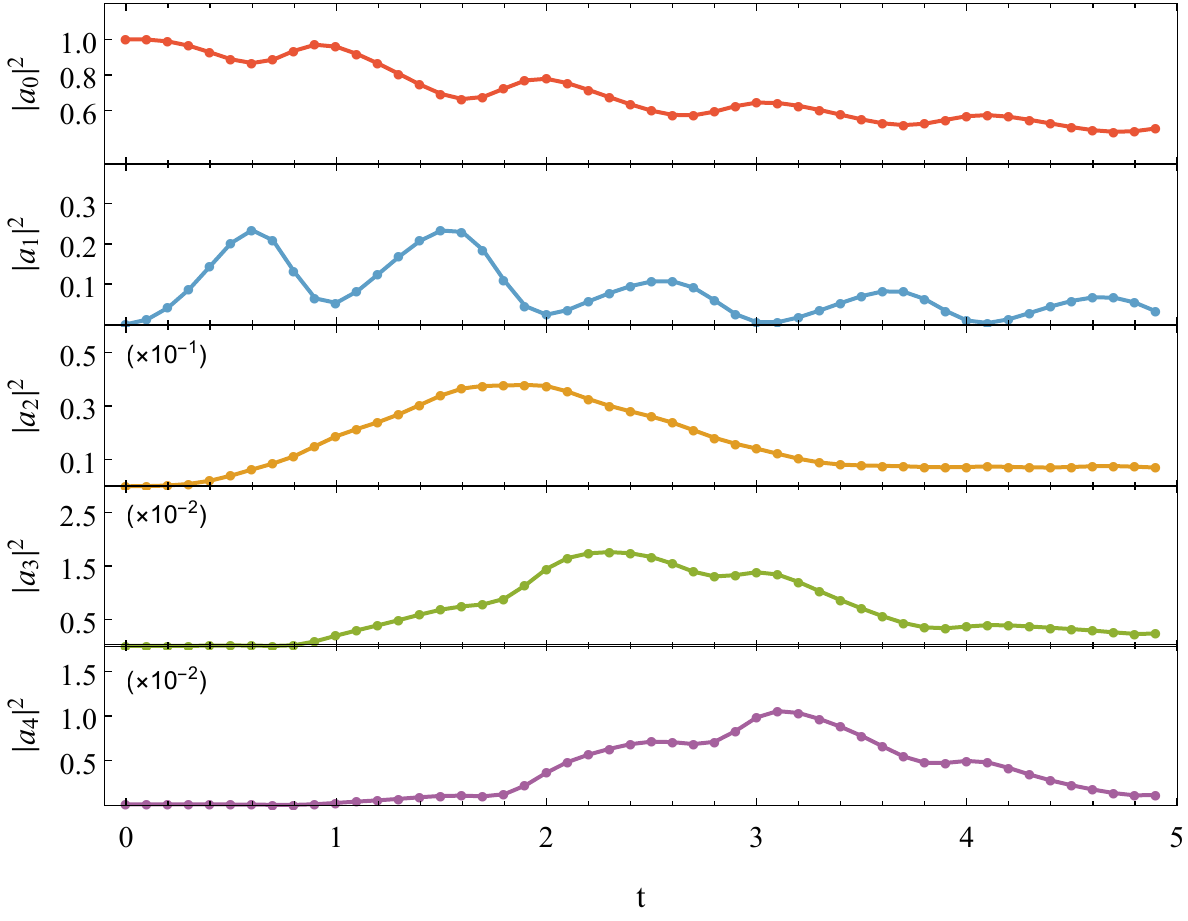}
\caption{Number of particles on site $j$ at different times. The trapping potential is $V=-7$, and the interaction strength is fixed as $U=1$. Typically, the drive strength is $g=5$ with drive frequency $\omega=6$. Note the scaled axes.}
\label{ajpic}
\end{figure}

Finally, in Fig.~\ref{ajpic} we visualize the ``jets" by plotting the number of particles on different sites as a function of time, which are calculated from Eq.~(\ref{ajt}).  The behavior on the sites near the origin are somewhat complicated, as they reflect the oscillations of the trapped condensate.  Further out ($j\geq 3$ for these parameters), one sees simpler behavior.
A clear wavefront is visible, moving at roughly a constant speed.
%\begin{equation}
%a_j(t)=\int^t d\tau (-J a_0(\tau)) G_{j1}(t-\tau).
%\end{equation}

\section{Summary and Outlook}
\label{summary}
Cold atom experiments have given us access to new classes of quantum dynamical systems.  In Ref.~\cite{fireworks1}, the experimentalists investigated the response of a Bose condensate to time-dependent interactions, and observed the emission of paired matter-wave jets. We explore the emission of both single-particle and pair jets with a simple lattice model where all of the physics is accessible.  

We analyze the emission processes within linear response, then present a detailed numerical study of nonlinear single-particle jets.
In our model, where the atoms are on a lattice, emission occurs when $|n \omega-|\nu||<2  J$.  In linear response single-particle emission corresponds to $n=1$ and pair emission to $n=1/2$.  Nonlinear single-particle emission corresponds to integer $n>1$.  Although we do not model it, nonlinear pair emission will occur when $n=m/2$ for integer $m>1$.
For deep traps, where the magnitude of the chemical potential  is large  $|\nu|\gg 2J$, the different emission channels are spectrally separated, and can be treated independently.  

The experiments in Ref.~\cite{fireworks1} differ in several ways from our model.  First, the atoms are not trapped on a lattice, and hence the excitation spectrum is unbounded, so instead of requiring $ |\nu|-2J<n\omega<|\nu|+2J$,   the $n$'th order excitations can occur whenever $n\omega>|\nu|$.  Furthermore, the experiment has a shallow trap, and $|\nu|$ is very small.  Consequently, all processes compete with one-another, and only the dominant pair emission processes (and their harmonics \cite{fireworks4}) are observed.  To see single-particle emission, one  would need to repeat the experiment with a deeper trap, where the modes are spectrally separated.

A related feature of the experiment which is not captured by our model, is that the experimental condensate contains many degrees of freedom. The pair emission process is accompanied by a dynamical instability of the condensate, and pattern formation \cite{fireworks2, fireworks3}.  This pattern formation is one of the reasons why pair emission dominates in the experiments \cite{yan, zhai, holland}.

%The simple model that we use cannot capture this pair emission, but as long as that instability is spectrally separated from the single-particle instability, it can be neglected.  Thus our theory for deep traps (large $V$).

% To gain understanding of these experiments we explore a simple 1D model.  Previous theoretical works have already explored the 

\section*{Acknowledgments}
This work was supported by the NSF Grant No. PHY-2110250. L.Q.L. received support from the China Scholarship Council (Grant No. 201906130092).

\begin{appendix}

\section{Derivation of the Green's function}
\label{Green's function}
Here we give a brief derivation of the Green's function for a semi-infinite chain of sites:  
\begin{equation}
i\partial_t G_{xy}(t)+ J G_{x,y+1}(t) + J G_{x,y-1}(t)= \delta_{xy} \delta(t),
\end{equation}
where $x,y=1,2,\cdots$, $G_{x0}\equiv 0$, and $G_{xy}(t<0)=0$.  This can be expressed in terms of the eigenstates of the homogeneous equations, $\psi_x^{(k)}=\sin(k x)$, where $0<k<\pi$, and $\epsilon_k=-2 J \cos k$.  By the orthogonality condition on the eigenstates, one can write
\begin{equation}
G_{xy}(t)=\frac{1}{i}\theta(t) \frac{2}{\pi}\int_0^\pi dk\, \psi_x^{(k)} \psi_y^{(k)} e^{-i\epsilon_k t},
\end{equation}
as this solves the homogeneous equation for $t>0$ and satisfies 
 $G_{xy}(t=0^+)=\delta_{xy}/i$.  Explicitly calculating the integrals yields
\begin{equation}
G_{xy}{(t)}= \frac{1}{i}\theta(t) \left(
i^{x-y} J_{x-y}(2 J t) - i^{x+y} J_{x+y}(2 J t)
\right),
\end{equation}
where $J_n(z)$ is the Bessel function of the first kind.
This expression has a natural interpretation in terms of images:  The first term represents direct propagation from $x$ to $y$, and the second term represents propagation of a fictitious image from $-x$ to $y$.

In the main text we require $G_{j1}(t)$.  Using the identities $J_{-n}(z)=(-1)^n J_n(z)$ and $J_{n-1}(z)+J_{n+1}(z)= (2n/z) J_n(z)$, we arrive at
\begin{equation}
G_{j1}(t)=i^{j-2} \frac{j J_j(2 J t)}{J t} \theta(t).
\end{equation}
We also need these Green's functions in the frequency domain.
The Fourier transforms obey $\omega G_{11}+J G_{21}=1$ and $\omega G_{j,1}+J G_{j-1,1}+J G_{j+1,1}=0$ for $j>1$.  This recursion relationship is solved by making the ansatz $G_{j1}= -e^{-\kappa j}/J$.  Solving the resulting quadratic equation for $e^{-\kappa}$ yields
\begin{equation}\label{freqdomain}
J G_{11}=z-i \sqrt{1-z^2},
\end{equation}
where $z=\omega/2J$.  The retarded function corresponds to adding an infinitesimal positive imaginary part to $z$, and taking the principle branch of the square root.  An alternative way to write this function is
$J G_{11}=z(1-\sqrt{1-z^{-2}})$. This latter representation is convenient, as when one takes the principle branch of the square root, the branch cut runs from $z=-1$ to $z=1$.

\section{Method of Multiple Scales}
\label{multiscale}
Here we reproduce
the single-particle emission results from Sec.~\ref{linear} by directly analyzing the lattice Gross-Pittaevskii equation,
Eq.~(\ref{mastereqn}), in the limit where $g(t)=g\sin(\omega t)$ is small.  

We use the method of multiple scales, writing
\begin{equation}
a_0(t) = e^{-i\nu t} \left(\alpha(t) +g u(t) e^{-i \omega t} +g v^*(t) e^{i\omega t} \right),
\end{equation}
where we take $\alpha(t),u(t)$, and $v(t)$ to be slowly varying.  In particular, the time derivatives of each of these quantities is suppressed by a factor of $g^2$.  
%Consequently 
%the loss rate $\Gamma = (-\partial_t |\alpha|^2)/|\alpha|^2$ scales as $g^2$.  %To this order we find that $\Gamma=0$ except when 
% $|\nu|-2J<\omega<|\nu|+2 J$.
 %that the loss rate $\Gamma = 2 \alpha^\prime/\alpha$ scales as $g^2$.  Outside of this window $\Gamma$ vanishes to this order.

We substitute this ansatz into Eq.~(\ref{mastereqn}), and collect terms which are linear in $g$.  The time derivatives of $u$ and $v$ do not appear at this order.  We neglect the slow variation of $u$ in the integral, writing
\begin{equation}
\int^t G_{11}(t-\tau) u(\tau) e^{-i (\omega+\nu)\tau}\,d\tau
\approx u(t) e^{-i(\omega+\nu)t} G_{11}(\omega+\nu).
\end{equation}
Equivalent expressions hold for $v$.  
Collecting the terms proportional to $e^{-i(\nu\pm\omega)t}$, yields
\begin{align}\label{ling}
\left(\begin{array}{cc}
\omega+\nu-\Delta_{\omega+\nu}& -\alpha^2 U\\
-(\alpha^*)^2 U& -\omega+\nu -\Delta^*_{-\omega+\nu}
\end{array}\right)
\left(\begin{array}{c}
u\\v
\end{array}
\right)  \nonumber \\
\quad= \frac{i |\alpha|^2}{2} \left(
\begin{array}{c}
\alpha\\\alpha^*
\end{array}
\right),
\end{align}
where, as in the main text
\begin{equation}
\Delta_\epsilon=V+2 |\alpha|^2 U+J^2 G_{11}(\epsilon).
\end{equation}
The matrix on the left of Eq.~(\ref{ling}) is simply the inverse of the Green's function in Eq.~(\ref{eom}), which can be easily inverted to find
\begin{equation}
\left(\begin{array}{c}
u\\v
\end{array}\right)= \frac{i |\alpha|^2}{2} {\cal G}
\left(\begin{array}{c}
\alpha\\\alpha^*
\end{array}\right).
\end{equation}
We then calculate the current,
% \begin{equation}
% i\partial_t\alpha = i g^2 \alpha \left(
% 2\alpha^* u-2 \alpha^* v+\alpha v - \alpha u
% \right)
% \end{equation}
% and hence
\begin{equation}
\langle \hat J_1 \rangle =-\partial_t |a_0(t)|^2
= g^2 |\alpha|^2 {\rm Re}(\alpha^* u-\alpha v),
\end{equation}
which agrees with Sec.~\ref{linear}.  In particular, $\Gamma=\langle \hat J_1 \rangle/|\alpha|^2$ is zero unless $-2J<\omega \pm \nu<2 J$.

\section{Simple Limit}\label{U=0}
Here we give explicit results for $\langle \hat J_1\rangle$ and $\langle \hat J_2\rangle$ when $U=0$.  This corresponds to oscillating the interaction strength about zero.  While the arithmetic is simpler, all aspects of the physics are still observed.

When $U=0$, the chemical potential is $\nu=J^2/V+V$, and stability requires $|V|>J$.  The off-diagonal elements of the matrix Green's function vanish, i.e., ${\cal G}_{12}={\cal G}_{21}=0$.  In frequency space the diagonal elements can be expressed as
\begin{eqnarray}
{\cal G}_{11}&=&\frac{-1}{V\omega}\left(\frac{\omega+\nu}{2}-V-i\sqrt{J^2-\frac{(\omega+\nu)^2}{4}}
\right), \\
{\cal G}_{22}&=&
\frac{1}{V\omega}\left(\frac{\nu-\omega}{2}-V+i\sqrt{J^2-\frac{(\nu-\omega)^2}{4}}
\right),
\end{eqnarray}
with spectral densities
\begin{eqnarray}\label{sd11}
A_{11}(\omega)&=&2 \sqrt{J^2-(\omega+\nu)^2/4}/(V\omega), \\
A_{22}(\omega)&=&2\sqrt{J^2-(\nu-\omega)^2/4}/(V\omega) \nonumber \\
&=& -A_{11}(-\omega),
\end{eqnarray}
when the arguments of the square roots are positive, and zero otherwise.
% $|\omega+\nu|<2 J$.  Similarly $A_{22}=??$ when $|\omega-\nu|<2J$. 
The single-particle excitation rate is then $\langle \hat J_1\rangle= (g^2\alpha^6/8 )\left[A_{22}(\omega)+A_{22}(-\omega)-A_{11}(\omega)-A_{11}(-\omega)\right]$.  When $|\omega+\nu|<2J$ it becomes
\begin{equation}
\langle \hat J_1\rangle =-\frac{g^2\alpha^6}{2} \frac{\sqrt{J^2-(\omega+\nu)^2/4}}{V\omega}.
\end{equation}
% {\color{blue}
% \begin{equation}
% \langle \hat J_1\rangle =-\frac{g^2\alpha^6}{2} \frac{\sqrt{J^2-(\omega+\nu)^2/4}-\sqrt{J^2-(\nu-\omega)^2/4}}{V\omega}.
% \end{equation}
% }
The pair excitation rate is $\langle \hat J_2\rangle
= (g^2/8) \left[\Pi(\omega)+\Pi(-\omega)\right]$, where $\Pi(\omega)=2{\rm Im} \chi(\omega)=2 \left[\chi^>(\omega)-\chi^<(\omega)\right]$, and the correlation functions are given by Eq.~(\ref{pairchi}).  When $U=0$ the ground state is a vacuum of the $\hat b$ operators, and hence ${\cal G}_{22}^>(t)=\langle \hat b_0^\dagger(0)\hat b_0(t)\rangle=0$.  Similarly, ${\cal G}_{11}^<$ and all the off-diagonal terms ${\cal G}_{12}^>, {\cal G}_{12}^<, {\cal G}_{21}^>, {\cal G}_{21}^<$ vanish.
%we have {\color{red} ${\cal G}_{11}^<={\cal G}_{22}^>=0$}{\color{blue} (feel confused)}, and all off-diagonal elements vanish. 
Therefore
\begin{eqnarray}
\chi^>(\omega)&=& -\alpha^4 \int \frac{dz}{2\pi}{\cal G}_{11}^>(\omega-z) {\cal G}_{11}^>(z),
\end{eqnarray}
and $\chi^>(\omega)$ vanishes if $\omega<0$.  Assuming $\nu<0$, we use ${\cal G}^>(\omega)=A(\omega)\theta(\omega)$ and  Eq.~(\ref{sd11}) to find
that $\chi^>(\omega)$ is nonzero when $-\nu-2 J<2\omega<-2\nu+2 J$, in which case
\begin{equation}
\chi^>(\omega)= \frac{4 \alpha^4}{-V^2} \int \frac{dz}{2\pi} \frac{\sqrt{\left(J^2-\frac{(\omega+\nu-z)^2}{4}\right)\left(J^2-\frac{(z+\nu)^2}{4}\right)}}{z(\omega-z)},
\end{equation}
where the integral is taken over $z$ such that all of the following inequalities are satisfied: $|\omega+\nu-z|<2J$, $|z+\nu|<2J$, and $0<z<\omega$.  
In particular,  if $-2 \nu-4 J<\omega<-2\nu $, then the integral runs from $z_-=-\nu-2 J$ to $z_+=\omega+\nu+2 J$.  Conversely, if $-2\nu<\omega<-2\nu+4 J$ the integral runs from $z_-=\omega+\nu-2 J$ to $z_+=-\nu+2 J$.  
The resulting integral can be expressed in terms of elliptic functions, though it is more efficient to simply evaluate the integral numerically. Similarly,
\begin{eqnarray}
\chi^<(\omega) &=& 
\alpha^4 \int \frac{dz}{2\pi} {\cal G}_{22}^<(\omega-z) {\cal G}_{22}^<(z) \nonumber \\
&=& -\chi^>(-\omega),
\end{eqnarray}
and
\begin{equation}
\langle \hat J_2\rangle = (g^2/2)\left[\chi^>(\omega)+\chi^>(-\omega)\right].
\end{equation}

\end{appendix}

\end{document}